\RequirePackage{fix-cm}
\documentclass[smallextended]{svjour3}       
\smartqed  
\usepackage{braket}
\usepackage{pgfplots}
\usepackage{csquotes}
\usepackage{float}
\usepackage{multirow}
\usepackage{dcolumn}
\usepackage[colorlinks=true, citecolor=blue]{hyperref}
\usepackage{hhline}
\usepackage{amssymb}
\usepackage[export]{adjustbox}
\usepackage{amsmath}

\usepackage{graphicx}

\usepackage{bm}
\usepackage{subcaption}
\usepackage{nameref}

\usepackage[none]{hyphenat}

\begin{document}
\sloppy 
\title{Experimental Demonstration of Non-local Controlled-Unitary Quantum Gates Using a Five-qubit Quantum Computer
}

\titlerunning{Experimental Demonstration of Non-local Controlled-Unitary Quantum Gates...}        

\author{Vishnu P. K. \and
        Dintomon Joy \and Bikash K. Behera \and Prasanta K. Panigrahi 
}


\institute{Vishnu P. K. \at Department of Physics, School of Physical Sciences, Central University of Kerala, Kasaragod 671314, Kerala, India
\\
\email{vishnupk07@gmail.com}
\and
Dintomon Joy \at Department of Physics, Cochin University of Science and Technology, Kochi 682022, Kerala, India\\
\email{dintomonjoy@cusat.ac.in}
\and
Bikash K. Behera \at
              Department of Physical Sciences, Indian Institute of Science Education and Research Kolkata, Mohanpur 741246, West Bengal, India\\
            \email{bkb13ms061@iiserkol.ac.in}           
           \and
              Prasanta K. Panigrahi \at   Department of Physical Sciences, Indian Institute of Science Education and Research Kolkata, Mohanpur 741246, West Bengal, India \\ \email{pprasanta@iiserkol.ac.in} 
}

\date{Received: date / Accepted: date}

\maketitle

\begin{abstract}
Local implementation of non-local quantum gates is necessary in a distributed quantum computer. Here, we demonstrate the non-local implementation of controlled-unitary quantum gates proposed by Eisert \emph{et al.} (Phys. Rev. A \textbf{62:052317}, 2000) using the five-qubit IBM quantum computer. We verify the fidelity and accuracy of the implementation through the techniques of quantum state and process tomographies.  
\end{abstract}


\section{Introduction}

A quantum computer can be efficiently used to solve daunting problems that exist in optimization \cite{KJIEEE,BassQST}, machine learning \cite{BiamonteNature2017}, artificial intelligence \cite{DunjkoRPP2018}, pattern recognition \cite{HolikFS2018}, cyber-security \cite{KaniaIEEE2017} to name a few. However, there are tremendous technological challenges to build a universal quantum computer \cite{Lagana}. As the number of qubits in a quantum computer increases, the effect of decoherence and the architectural complexity introduces certain limitations in controlling and manipulating the fragile quantum information. Hence, instead of building a quantum processor with large number of qubits, a multiprocessor device with smaller number of qubits is the best way to construct. A large network of small quantum computers connected by entanglement and classical communication channels, gives rise to a distributed quantum computer where each quantum processor acts as a node. Such distributed quantum computation has been used to find profound application in the phase estimation problem \cite{CiracPRA1999}. In a distributed quantum computer, it is necessary to have optimal implementation of quantum gates between qubits that are placed in different quantum processors. This problem has been addressed and optimal protocols have been proposed by Eisert \emph{et al.} \cite{PhysRevA.62.052317}. In lattice model \cite{Gavin} of quantum computers, qubits are fixed in position and the interaction happens between nearest qubits. Therefore to connect distant qubits, the states of these qubits must be swapped until they become adjacent to each other. Then requisite operation can be performed and the states can be swapped back to their initial positions. This method introduces too much overhead in the number of CNOT gates required, as three CNOT gates are needed to implement one SWAP gate, which increases linearly with the distance between the two qubits. One may also consider teleportation as an option to perform a non-local gate between qubits in two spatially separated nodes (quantum processors). In this case, the quantum state of one of the qubits gets teleported to the distant node, the required operation between the two qubit states are performed there and then teleported back to the original node. This method requires 2 ebits (two Bell pairs), 4 cbits (classical information) and some unitary operations to reconstruct the state. Though this `two-way' teleportation technique seems effective, the resource requirement can be reduced further by following the protocol proposed by Eisert \emph{et al.} \cite{PhysRevA.62.052317}, which allows non-local gates like Controlled-NOT to be implemented between two spatially separated qubits with one maximally-entangled Bell state and one bit of classical information in either direction as physical resources. 

This opens a way for an efficient distributed quantum computation or quantum network among the distributed users and also facilitates long distance quantum information processing. There are many experimental studies which demonstrate the non-local implementation of quantum gates \cite{PhysRevLett.90.117901,PhysRevLett.93.240501,PhysRevA.72.032333,1464-4266-7-10-026,PhysRevA.71.064302,PhysRevLett.94.030501,PhysRevA.71.044304,PhysRevLett.96.010503,PhysRevA.76.044305,PhysRevA.87.062337,PhysRevA.90.012311}. Yimsiriwattana and Lomonaco \cite{quant-ph/0403146} had also proposed a distributed implementation of Shor's quantum factoring algorithm on a distributed quantum network model using entanglement and non-local operations.
 
IBM (International Business Machines Corporation) has attracted the global scientific community by developing a five-qubit quantum computer made up of superconducting charge-qubits called transmon qubits \cite{PhysRevA.76.042319}. It provides a free access through a cloud based web-interface called IBM Quantum Experience (IBM QE) \cite{research}, which allows researchers to design, test and run their experiments. Various theoretical protocols \cite{1607.02398,PhysRevA.94.012314,PhysRevA.95.032131,PhysRevA.94.032329,gubaidullina2017stability,PhysRevA.95.052339,1610.06980,Linke28032017,MS2,MS3,MS1,AK1,BKB2,BKB3,BKB4,1706.04341,1706.08080} have already been tested and verified in this platform. Here the non-local implementation of controlled-unitary quantum gates proposed by Eisert \emph{et al.} \cite{PhysRevA.62.052317} has been explicated using the IBM quantum computer.

The paper is organized as follows. Sec. \ref{II} gives an outline of the implementation of non-local quantum gates proposed by Eisert \emph{et al.} \cite{PhysRevA.62.052317}. Sec. \ref{III} demonstrates the experimental realization of the above protocol using IBM's 5 qubit quantum computer. Following which, Sec. \ref{IV} characterizes our implementation by the method of quantum state and process tomographies. Finally we conclude in Sec. \ref{V}. 

\section{Non Local Implementation of Controlled-Unitary Gates \label{II}}
  
The protocol proposed by Eisert \emph{et al.} \cite{PhysRevA.62.052317} for the implementation of non-local gates between spatially separated qubits, is described as follows. Here, one of the users (Alice) wants to operate a controlled unitary gate on another qubit which is located far away in possession of another user (Bob) using minimal physical resources. Here, Alice's qubit acts as the control qubit, given by

\begin{equation}
\ket{\psi_{A}}=\alpha\ket{0}+\beta\ket{1}
\label{eq1}
\end{equation}
and Bob's qubit, which acts as 
a target qubit, is given by
\begin{equation}
\label{eq2}
\ket{\psi_{B}}=\gamma\ket{0}+\delta\ket{1}
\end{equation}

The following is the entanglement channel shared by Alice and Bob.
\begin{equation}
\ket{\phi^{+}}_{ab}=\frac{1}{\sqrt{2}}(\ket{00}+\ket{11})_{ab}
\end{equation}

The initial state of the system can be written as,
\begin{equation}
\begin{split}
\ket{\Psi}_{i}&=\ket{\psi_{A}}\otimes \ket{\phi^{+}}_{ab}\otimes \ket{\psi_{B}}\\
&=\frac{1}{\sqrt{2}}(\alpha\ket{000}+\alpha\ket{011}+\beta\ket{100}+\beta\ket{111})_{Aab} \otimes \ket{\psi_{B}}\\
\end{split}
\end{equation}

Now, Alice applies a CNOT operation between her unknown qubit $A$ and qubit $a$ in her possession. The resultant state of Alice's system is given by,
\begin{equation}
\ket{\psi_{Aab}}=\frac{1}{\sqrt{2}}(\alpha\ket{000}+\alpha\ket{011}+\beta\ket{110}+\beta\ket{101})
\end{equation}
In the next step, Alice measures her qubit $a$ in the computational basis $\{\ket{0},\ket{1}\}$ and conveys her results to Bob through a classical channel. Bob applies a Pauli-X gate on qubit $b$, if the classical message conveyed is `1' or otherwise applies identity (I) gate.
Then the state becomes, 
\begin{equation}
\ket{\psi}_{Ab}=\frac{1}{\sqrt{2}}(\alpha\ket{00}+\beta\ket{11})_{Ab}
\end{equation} The combined state of the whole system,
\begin{equation}
\begin{split}
\ket{\psi_{AbB}}=\frac{1}{\sqrt{2}}(\alpha\ket{00}+\beta\ket{11})_{Ab}\otimes \ket{\psi_{B}}
\end{split}
\end{equation}
Now, Bob performs the required controlled-unitary ($\hat U$) operation on the target qubit ($B$) with qubit ($b$) acting as the control qubit. This leads to the state given by
\begin{equation}
\ket{\psi_{AbB}}=\frac{1}{\sqrt{2}}(\alpha\ket{00}\ket{\psi_{B}}+\beta\ket{11}\hat{U}\ket{\psi_{B}})
\end{equation}
In order to disentangle the ancillary qubit ($b$) from $A$ and $B$, a Hadamard gate is applied on it resulting in the state, 
\begin{equation}
\ket{\psi_{AbB}'}=\frac{1}{2}(\alpha\ket{00}\ket{\psi_{B}}+\alpha\ket{01}\ket{\psi_{B}}+\beta\ket{10}\hat{U}\ket{\psi_{B}}-\beta\ket{11}\hat{U}\ket{\psi_{B}})\\
\end{equation}
Finally, Bob measures $b$ in the computational basis and the result is conveyed to Alice classically. If the conveyed result is `1', Alice performs Pauli-Z operation on her unknown qubit or otherwise an identity operation is performed as given in Table \ref{tab1}. This leads to the final state,
\begin{equation}
    \ket{\Psi}_{f}
    =\alpha\ket{0}\ket{\psi_{B}}+\beta\ket{1}\hat{U}\ket{\psi_{B}}
\end{equation}
which is the required ouput state of a controlled-unitary operation.

\begin{table}

\centering
\begin{tabular}{|c|c|}
\hline
Result of Bob's measurement& Operation performed\\
\hline
 $\ket{0}$ & I \\
 $\ket{1}$ & Z\\
 \hline
\end{tabular}
\caption{Operations performed by Alice based on Bob's measurement result.}
\label{tab1}
\end{table}
 
\section{Implementation in IBM Quantum Computer \label{III}}

\subsection{Quantum Gates and Some Gate Combinations}

The single qubit quantum gates like Identity gate (I), Pauli gates (X, Y, Z), Hadamard gate (H) and phase gates (S, S$^{\dagger}$, T, T$^{\dagger}$) are available in the IBM quantum experience tool box. These gates can be inserted anywhere in the circuit using a graphical user interface which allows click, drag and drop method. The output state of each line can be obtained by placing a measurement operator at the end of these lines, which in turn gives the output state in the computational basis $\{\ket{0}, \ket{1}\}$ along with its probabilities. 
\begin{figure}
   \centering
   \includegraphics[scale=0.6]{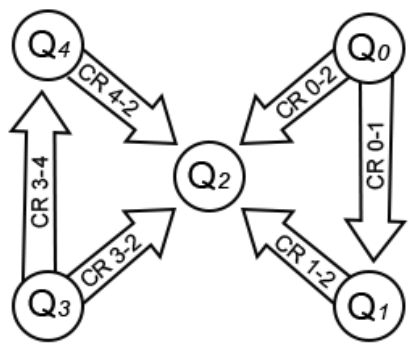}
    \caption{IBM five-qubit quantum chip ibmqx2's architecture (Credits-IBM).}
    \label{fig:01}
\end{figure}

\begin{figure}
\raggedright
    \begin{subfigure}[b]{0.55\textwidth}
   \centering
   \includegraphics[scale=0.36]{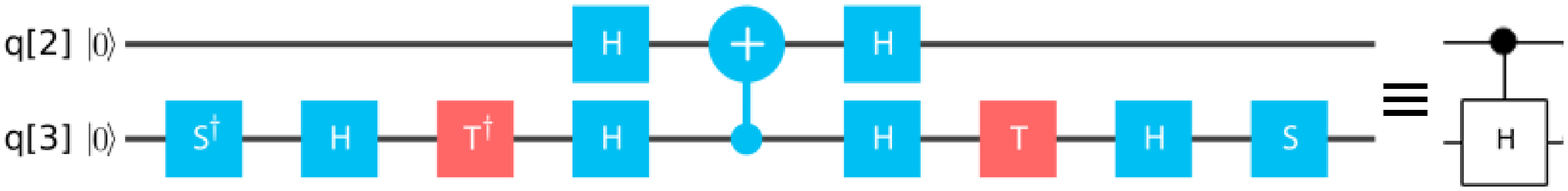}
    \caption{}
    \label{3a} 
    \end{subfigure}
    \begin{subfigure}[b]{0.55\textwidth}
   \centering
   \includegraphics[scale=0.36]{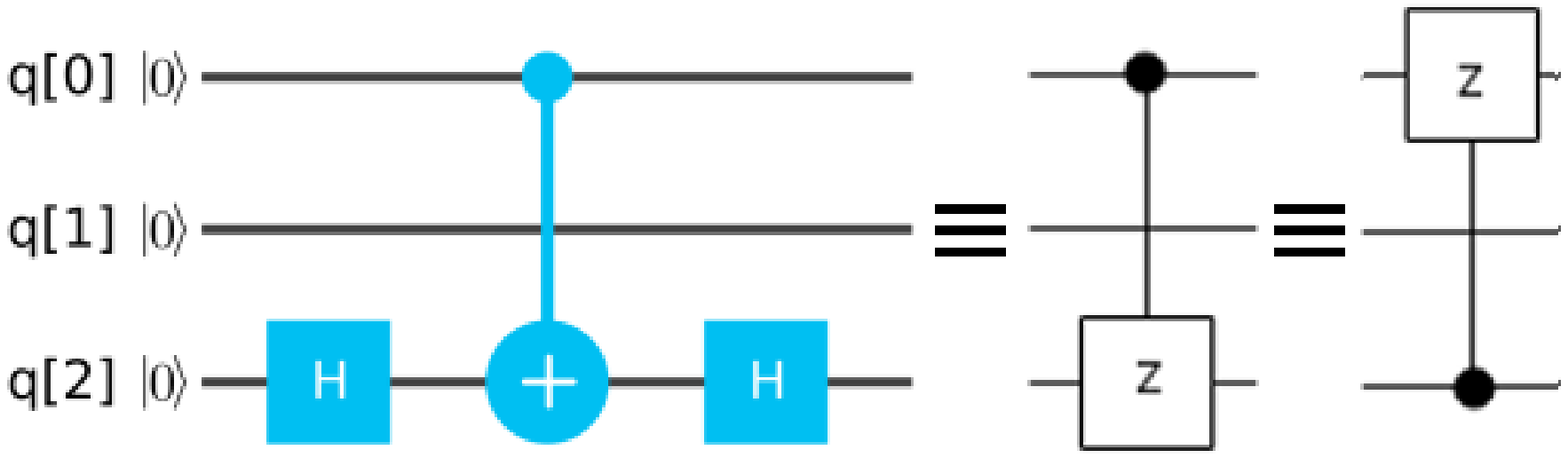}
    \caption{}
    \label{3b}
    \end{subfigure}
\caption{The figure depicts some relevant operations and their equivalent circuits: (a) Controlled-H operation from $Q_{2}$ to $Q_{3}$. (b) Controlled-Z operation from $Q_{3}$ to $Q_{0}$.}
\label{fig:02}
\end{figure}

Unlike single qubit gates, the architecture of of IBM 5-qubit quantum processor (Fig. \ref{fig:01}) needs to be considered to implement a two qubit operation like CNOT. Here $Q_{0}$, $Q_{1}$, $Q_{2}$, $Q_{3}$ and $Q_{4}$ represent the five qubits of the quantum processor `ibmqx2' on which any quantum gate can be implemented. The direction of arrows depicts the direction of CNOT gate to be implemented between two qubits. Qubits ($Q_1, Q_2, Q_4$) having arrow pointed towards them, are used as the target qubits and the qubits ($Q_0, Q_1, Q_3, Q_4$) having arrows away from them, act as the control qubits. The CNOT gate can be applied between any two qubits in any order by following the protocols given in Ref. \cite{BKB1}. Other two qubit gates can also be implemented by using equivalent circuits as given in Fig. \ref{fig:02}. It is suggested that, in cases where the qubits are not directly connected by an arrow, Eisert's scheme \cite{PhysRevA.62.052317} may be used to perform non-local operation rather than swapping the states. Because, swapping requires more number of gates than Eisert's scheme when implemented in the IBM processor and may induce more errors and decoherence in the output. In our experiment, the following equivalent circuits have been used for Controlled-H (CH) gate and Controlled-Z (CZ) gate (Fig. \ref{fig:02}).   

In IBM quantum experience, the real experiments have been carried out using the five-qubit quantum processor, `ibmqx2' by choosing different number of shots, e.g., 1024, 4096 and 8192. Here, shots represent the number of times a given experiment is run in a quantum processor. With single run we will not get any sensible or useful result from the experiment. The quantum processor runs the experiment several times (based on the no. of shots given) and then displays the results in the form of a histogram. In this histogram, each bar represents the ‘probability’ of getting one of the possible results for that experiment. The results will become much more accurate as we run the experiment several times. Shots - 8192, is the maximum number of times we can run an experiment in the quantum processor and therefore it gives the accurate result of an experiment. The `Custom Topology' allows to design and classically simulate the results of quantum circuits up to 20 qubits taking any number of shots starting from 1 to 8192. The interface provides details about the possible gate errors after the execution of single qubit and two-qubit gates on the real-chip. More description of the device characteristics and data analysis are available in Refs. \cite{1706.04341,1703.10793}. 

\subsection{Non-local CNOT Gate}
In this section, the non-local implementation of CNOT gate between arbitrary qubit states of Alice ($Q_0$) and Bob ($Q_3$) has been illustrated. Two different arbitrary unknown initial states have been chosen for Alice and Bob. The state of Alice's qubit is given as, 
\begin{equation}
\begin{split}
\ket{\psi_{A}}=HTH\ket{0}&=\frac{1}{2}((1+e^{i\frac{\pi}{4}})\ket{0}+(1-e^{i\frac{\pi}{4}})\ket{1})
\end{split}
\label{eq16}
\end{equation}

Whereas, the state of Bob's qubit is the following,
\begin{equation}\label{eq17}
\begin{split}
\ket{\psi_{B}}=HSTH\ket{0}&=\frac{1}{2}((1+ie^{i\frac{\pi}{4}})\ket{0}+(1-ie^{i\frac{\pi}{4}})\ket{1})
\end{split}
\end{equation}
The resultant initial state of both Alice and Bob is, 
\begin{equation}
\ket{\psi_{AB}}=\frac{i}{2\sqrt{2}}\ket{00}+\frac{(\sqrt{2}+1)}{2\sqrt{2}}\ket{01}+\frac{(\sqrt{2}-1)}{2\sqrt{2}}\ket{10}-\frac{i}{2\sqrt{2}}\ket{11}
\end{equation}

\begin{figure}[H]
   \centering
   \includegraphics[scale=0.36]{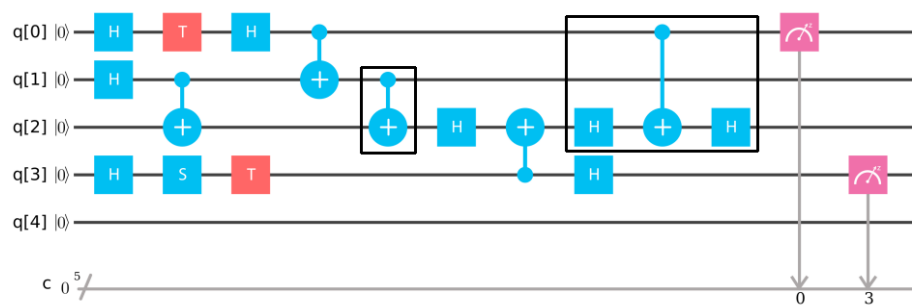}
    \caption{IBM quantum circuit illustrating the non-local implementation of a CNOT gate, where qubit q[0] is the control qubit of Alice and q[3] is the target qubit of Bob.}
    \label{fig:03}
\end{figure}

After non-local CNOT operation, the state becomes
\begin{equation}
\ket{\psi^{'}_{AB}}=\frac{i}{2\sqrt{2}}\ket{00}+\frac{(\sqrt{2}+1)}{2\sqrt{2}}\ket{01}-\frac{i}{2\sqrt{2}}\ket{10}+\frac{(\sqrt{2}-1)}{2\sqrt{2}}\ket{11}
\end{equation}

\begin{figure}[H]
   \centering
   \includegraphics[scale=0.36]{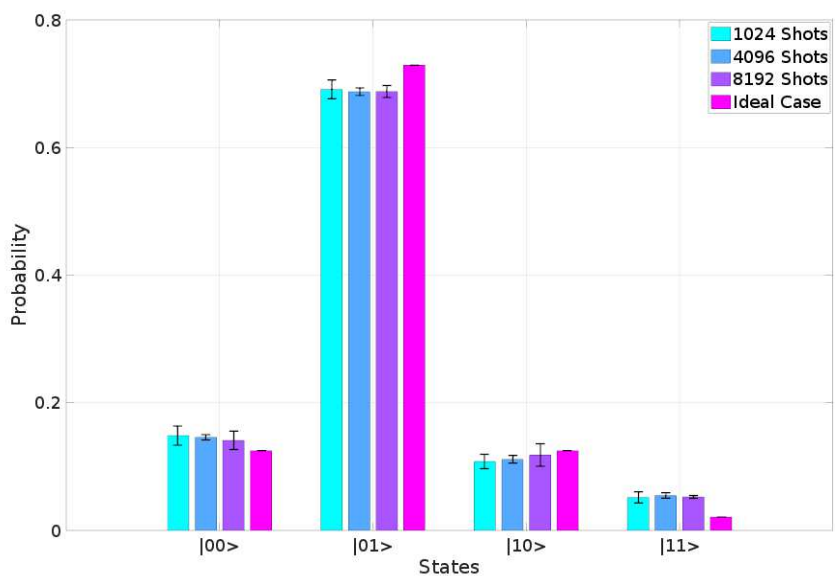}
    \caption{The histograms show Ideal and Run results of non-local CNOT operation for different number of shots.}\label{fig:04}
\end{figure}
The equivalent quantum circuit of Eisert's scheme (Fig. \ref{fig:03}) has been executed in IBM quantum computer. for three different sets of number of shots. The results and the corresponding bar diagrams have been presented Fig. \ref{fig:04} respectively. Each experiment has been performed for 10 times. The error bars represent the standard deviation of experimental data.
The statistical fidelity of the operation \cite{carol}, $F_{s}=\sum_{j=0}^{3}\sqrt{p_{j}^{exp}p_{j}^{th}}$, (where $p_{j}^{exp}$ and $p_{j}^{th}$ represent the respective experimental and theoretical probability of measurements for $j^{th}$ state) is found to be 
$F_{s}^{CN}=0.995\pm 0.002$.

\subsection{Non-local CH Gate}

\begin{figure}
   \centering
   \includegraphics[scale=0.23]{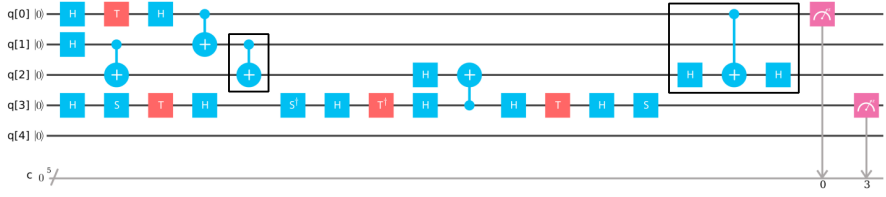}
    \caption{IBM quantum circuit illustrating the non-local implementation of Controlled-H gate.}
    \label{fig:05}
\end{figure}

The implementation of non-local CH gate (Fig. \ref{fig:05}) is similar to that of previous case, where the CNOT gate between q[2] and q[3] is replaced by an equivalent circuit \cite{6516700} of Controlled-Hadamard Gate (Fig. \ref{3a}). The same arbitrary states (Eqs. \eqref{eq16} and \eqref{eq17}) are also used for the initial qubits of Alice and Bob. The final state obtained after CH operation is given by,
\begin{equation}
\ket{\psi^{'}_{AB}}=\frac{i}{2\sqrt{2}}\ket{00}+\frac{(\sqrt{2}+1)}{2\sqrt{2}}\ket{01}+\frac{(\sqrt{2}-1-i)}{4}\ket{10}+\frac{(\sqrt{2}-1+i)}{4}\ket{11}
\end{equation}

\begin{figure}[H]
   \centering
   \includegraphics[scale=0.36]{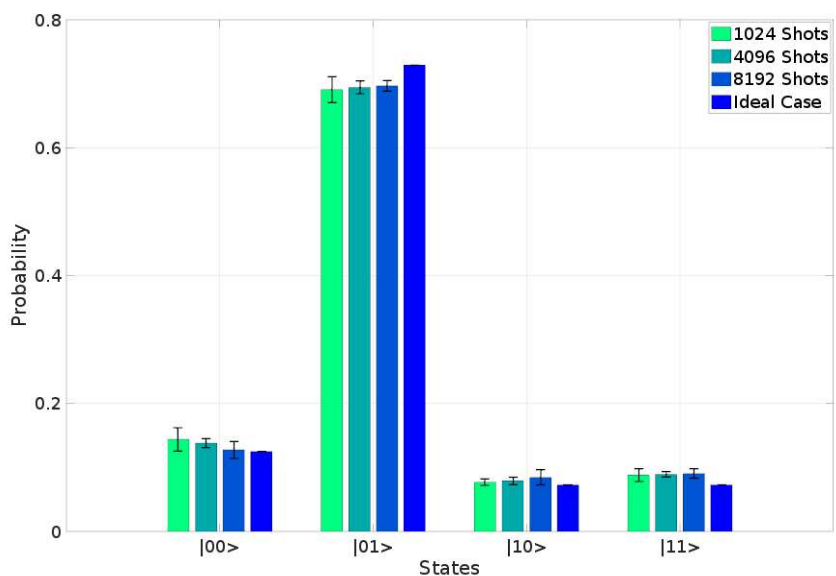}
    \caption{The histograms show Ideal and Run results of non-local CH for different number of shots. }
    \label{fig:06}
\end{figure}

The obtained results are given in Fig. \ref{fig:06}. The statistical fidelity of operation for this case is found to be $F_{s}^{CH}=0.998\pm 0.002$

The discrepancy in observed and theoretical results are due to several reasons like decoherence effects, gate errors, state preparation error, measurement errors etc. The device parameters of the quantum processor has been shown in the following table \ref{tab3}, which presents qubit readout error, gate error, relaxation time ($T_{1}$) and coherence time($T_{2}$) of the qubits. 

\begin{table}
\centering
\begin{tabular}{| c | c | c | c | c |}
 \hline
 Qubit  & Q0 & Q1 & Q2 & Q3 \\
 \hline
Gate Error($10^{-3}$) & 1.37 & 1.37 & 2.23 & 1.72  \\
Read Out Error($10^{-2}$) & 2.40 & 2.60 & 3.00 & 2.20 \\
$T_{1}$ ($\mu s$)  & 62.4 & 55.1 & 48.4 & 59\\
$T_{2}$ ($\mu s$) & 77.5 & 64 & 54.7 &  57.3  \\
f(GHz) &  5.276 & 5.212 & 5.015 &   5.280  \\
\hline
Multiqubit Gate Error &$CX_{01}$ 0.0272 &$CX_{02}$  0.0417 &$CX_{12}$ 0.0376 &$CX_{32}$ 0.0397\\
\hline
\end{tabular}
\caption{The table shows the device calibration parameters of ibmqx2. The Fridge Temperature is kept at 0.0159 K}
\label{tab3}
\end{table}

\section{Characterization Techniques \label{IV}}
\subsection{Quantum State Tomography}
Quantum state tomography is a well known method to characterize a quantum state \cite{MS2,MS3,PhysRevA.64.012317,PhysRevA.64.052312,tomo,Niggebaum}, which includes comparison of theoretical and experimental density matrices.

\begin{figure}[H]
\centering
    \begin{subfigure}[b]{0.55\textwidth}
   \centering
   \includegraphics[scale=0.45]{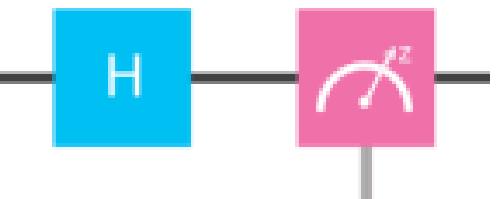}
    \caption{}
    \label{7a}
    \end{subfigure}
    \begin{subfigure}[b]{0.55\textwidth}
   \centering
   \includegraphics[scale=0.45]{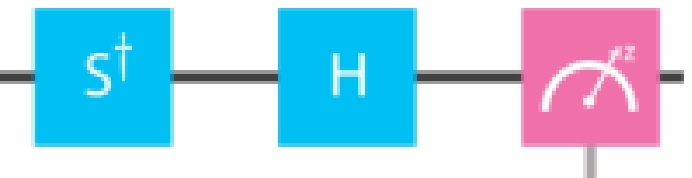}
    \caption{}
    \label{7b}
    \end{subfigure}
    \begin{subfigure}[b]{0.55\textwidth}
   \centering
   \includegraphics[scale=0.45]{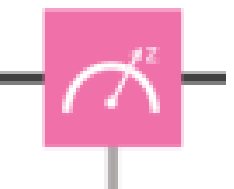}
    \caption{}
    \label{7c}
    \end{subfigure}
\caption{Figure depicting measurement schemes for different bases: (a) X, (b) Y and (c) Z.}
\label{fig:07}
\end{figure}

The theoretical density matrix of the initially prepared quantum state is given by,
\begin{equation}
\rho^{T}= \ket{\Psi}\bra{\Psi}    
\end{equation}
and the expression for the experimental density matrix of a multi-qubit system is given by the following expression.
\begin{equation}
\rho^{E}=\frac{1}{2^{N}}\sum_{i_{1},i_{2},i_{3}...i_{N}=0}^{3}T_{i_{1}i_{2}i_{3}...i_{N}}(\sigma_{i_{1}} \otimes \sigma_{i_{2}}\otimes\sigma_{i_{3}}...\sigma_{i_{N}})
\label{eq22}
\end{equation}
where $\sigma_{i_{N}}$ represents the Pauli matrices acting on the $N^{th}$ qubit. The quantity $T_{i_{1}i_{2}i_{3}...i_{N}}$ denotes the outcome of a specific projective measurement in the experiment, which is related to Stokes parameters in Bloch sphere \cite{PhysRevA.64.052312,tomo}. This eq. contains $4^{N}$ terms, however, only $3^{N}$ set of measurements are required, where each set has a particular combination of measurement basis. For a two qubit system, Eq. \eqref{eq22} reduces to
\begin{equation}
\rho^{E}=\frac{1}{4}\sum_{i_{1},i_{2}=0}^{3}T_{i_{1}i_{2}}(\sigma_{i_{1}} \otimes \sigma_{i_{2}})
\label{eq23}
\end{equation}
where
\begin{equation}
T_{i_{1}i_{2}}=S_{i_{1}} \times S_{i_{2}}
\end{equation}
where the indices $i_{1}$ and $i_{2}$ can take values 0, 1, 2 and 3 corresponding to I, X, Y and Z Pauli matrices respectively. For a single qubit case, the Stokes parameters are $S_{0}=P_{\ket{0}}+P_{\ket{1}}$,
$S_{1}=P_{\ket{0_{X}}}-P_{\ket{1_{X}}}$,
$S_{2}=P_{\ket{0_{Y}}}-P_{\ket{1_{Y}}}$,
$S_{3}=P_{\ket{0_{Z}}}-P_{\ket{1_{Z}}}$, where P represents the probability of success for the corresponding bases given in the subscript. For two-qubit system, the probability of measurement outcomes are determined by $P_{\ket{00}}$, $P_{\ket{01}}$, $P_{\ket{10}}$ and $P_{\ket{11}}$ in the appropriate basis. The different measurement bases can be prepared by operating proper gates before the measurement operation as shown in Fig. \ref{fig:07}. The expressions of $T_{i_{1}i_{2}}$ in terms of probability outcome has been provided in Table \ref{tab5}.    

\begin{table}[H]
\centering
\begin{tabular}{| c | c |}
 \hline

 $T_{i_{1}i_{2}}$  & Expansion in terms of Probabilities  \\
 \hline
$T_{00}$ & $P_{\ket{00}}$+$P_{\ket{01}}$+$P_{\ket{10}}$+$P_{\ket{11}}$ \\
$T_{0i_{2}}$ & $P_{\ket{00}}$-$P_{\ket{01}}$+$P_{\ket{10}}$-$P_{\ket{11}}$ \\
$T_{i_{1}0}$ & $P_{\ket{00}}$+$P_{\ket{01}}$-$P_{\ket{10}}$-$P_{\ket{11}}$ \\
$T_{i_{1}i_{2}}$ & $P_{\ket{00}}$-$P_{\ket{01}}$-$P_{\ket{10}}$+$P_{\ket{11}}$\\
\hline

\end{tabular}
\caption{The table depicts the expression for $T_{i_{1}i_{2}}$, where $i_{1}$, $i_{2}$ can take values 1, 2 and 3 according to X, Y, and Z bases respectively.}
\label{tab5}
\end{table}
\begin{figure}[t]
   \centering
   \includegraphics[scale=0.50]{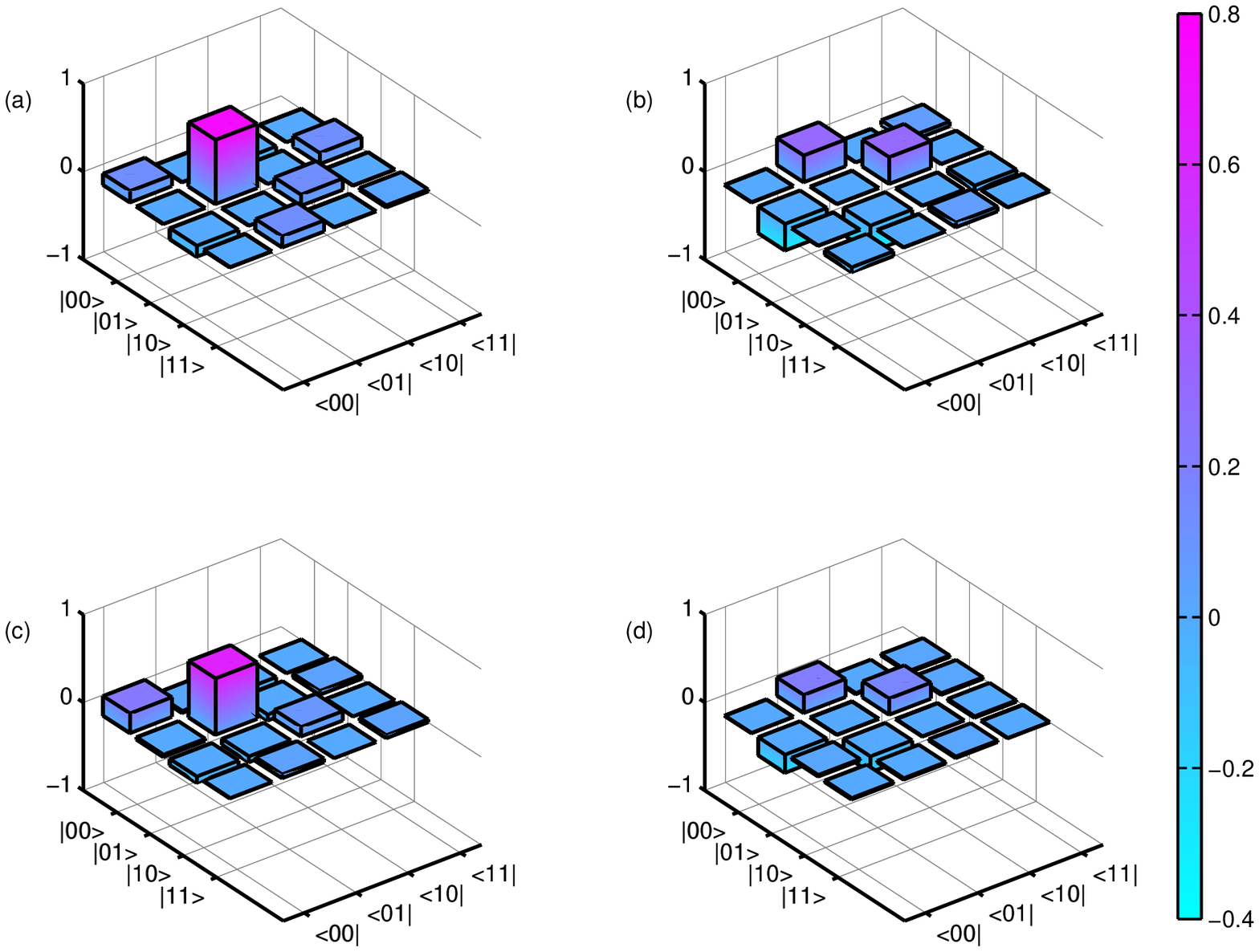}
    \caption{The figure depicts both the real and imaginary parts of ideal and experimental density matrices for non-local CNOT implementation. (a), (b): Ideal case; (c), (d): Experimental case.}\label{fig:08}
\end{figure}

\begin{figure}[t]
   \centering
   \includegraphics[scale=0.50]{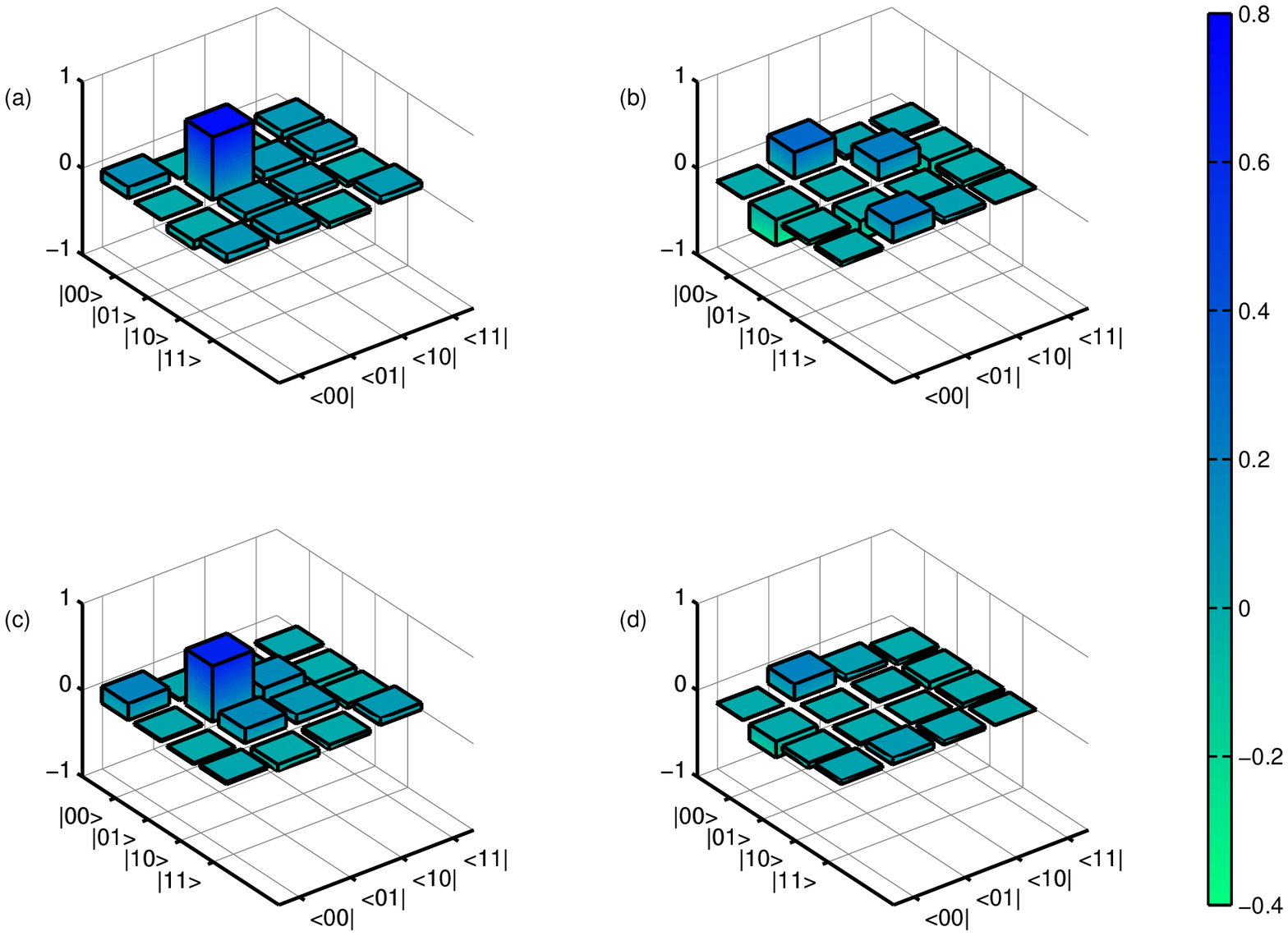}
    \caption{The figure depicts both the real and imaginary parts of ideal and experimental density matrices for non-local CH implementation. (a), (b): Ideal case; (c), (d): Experimental case.}
    \label{fig:09}
\end{figure}

The experimental density matrix (Eq. \eqref{eq23}) can be expanded as,
\begin{equation}
\begin{split}
\rho^{E}=&\frac{1}{4}[T_{II}(I\otimes I)+T_{IX}(I \otimes \sigma_{X})+T_{IY}(I \otimes \sigma_{Y})+T_{IZ}(I \otimes \sigma_{Z})\\&+T_{XI}(\sigma_{X} \otimes I)
+T_{XX}(\sigma_{X} \otimes \sigma_{X})+T_{XY}(\sigma_{X} \otimes \sigma_{Y})+T_{XZ}(\sigma_{X} \otimes \sigma_{Z})\\&+T_{YI}(\sigma_{Y} \otimes I)+T_{YX}(\sigma_{Y} \otimes \sigma_{X})+T_{YY}(\sigma_{Y} \otimes \sigma_{Y})+T_{YZ}(\sigma_{Y} \otimes \sigma_{Z})\\&+T_{ZI}(\sigma_{Z} \otimes I)+T_{ZX}(\sigma_{Z} \otimes \sigma_{X})+T_{ZY}(\sigma_{Z} \otimes \sigma_{Y})+T_{ZZ}(\sigma_{Z} \otimes \sigma_{Z})]
\end{split}
\end{equation}
The fidelity between ideal and prepared arbitrary states of qubits $A$ and $B$ is calculated from \cite{nil,mcm},

\begin{equation}
\begin{split}
F(\rho^{T},\rho^{E})&=Tr\left(\sqrt{\sqrt{\rho^{T}}\rho^{E}\sqrt{\rho^{T}}}\right)\\
&=Tr\left(\sqrt{\ket{\Psi}\bra{\Psi}\rho^{E}\ket{\Psi}\bra{\Psi}}\right)
\end{split}
\end{equation}
Fidelity measures the overlap between two density matrices and hence quantifies the closeness of theoretical and experimental quantum states obtained as output. From Fig. \ref{fig:08}, the accuracy can be easily checked by comparing the theoretical and experimental density matrices for non-local CNOT implementation. Fidelity of the experimental result is found to be $F^{CN}=0.879$. Similarly, Fig. \ref{fig:09} depicts the theoretical and experimental density matrices for non-local CH gate implementation. The fidelity of this experiment is calculated to be $F^{CH}=0.831$.

\subsection{Quantum Process Tomography}

In this technique, a process matrix is constructed to perform the complete characterization of applied quantum gate operations \cite{quant-ph/9610001,PhysRevLett.78.390,PhysRevA.64.012314,PhysRevLett.93.080502,cs/0012017,White:07,PhysRevA.80.042103,doi:10.1063/1.4774119,PhysRevB.90.144504}. Here, quantum process tomography of output states has been performed for all combinations of the particular set of input states. The input states for two-qubit system used are,  

\begin{equation}
    \ket{\text{H}}=\ket{0},
    \ket{\text{V}}=\ket{1},
    \ket{\text{D}}=\frac{\ket{0}+\ket{1}}{\sqrt{2}},
    \ket{\text{R}}=\frac{\ket{0}+i\ket{1}}{\sqrt{2}}
\end{equation}

If $\rho^{\alpha\beta}$ is input density matrix, corresponding output is represented by $\varepsilon(\rho^{\alpha\beta})$, where $\alpha,\beta$ $\in$  $\{\text{H,V,D,R}\}$.

\begin{figure}
  \begin{subfigure}{\linewidth}
  \adjincludegraphics[width=.5\linewidth]{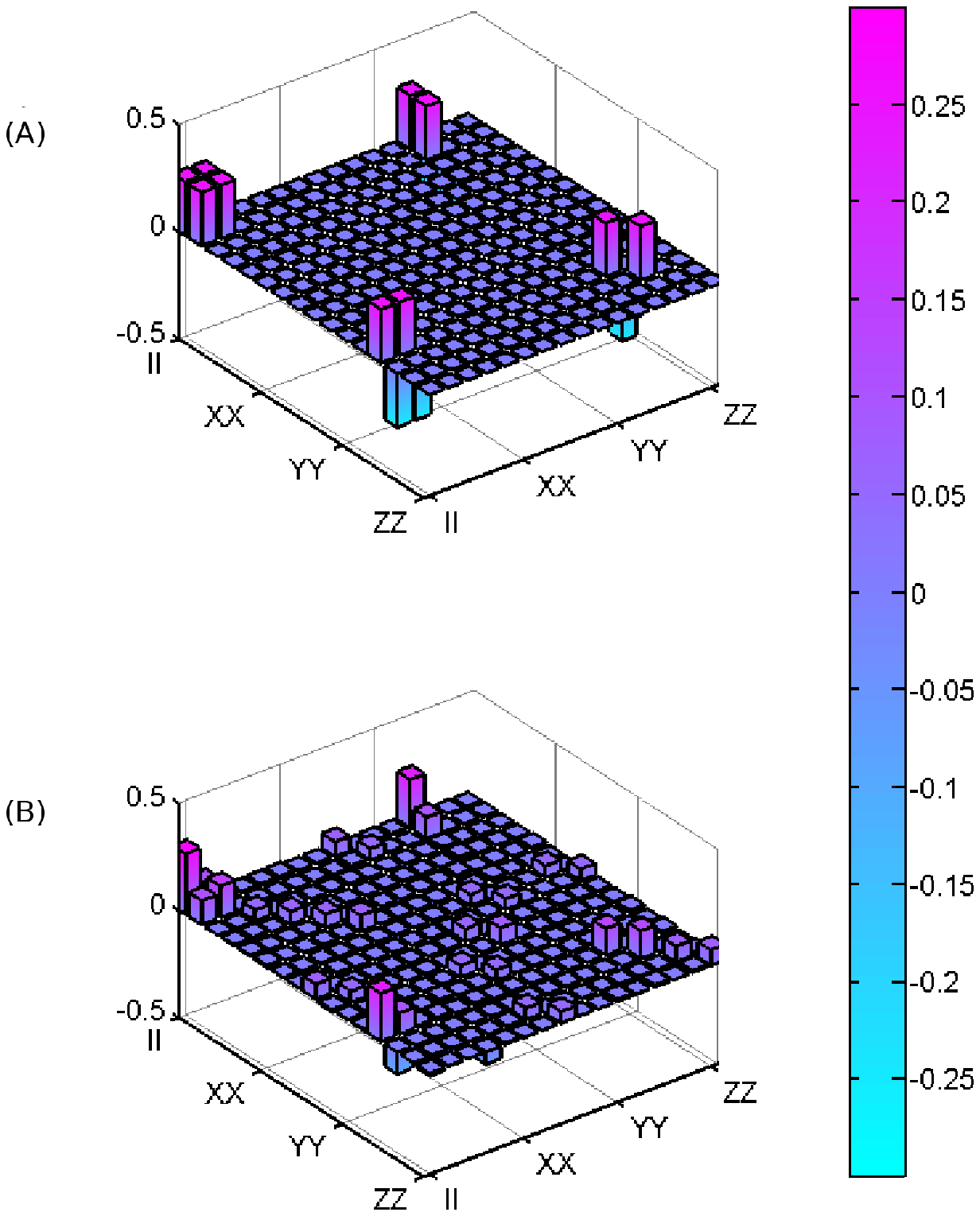}\hfill
  \adjincludegraphics[width=.5\linewidth]{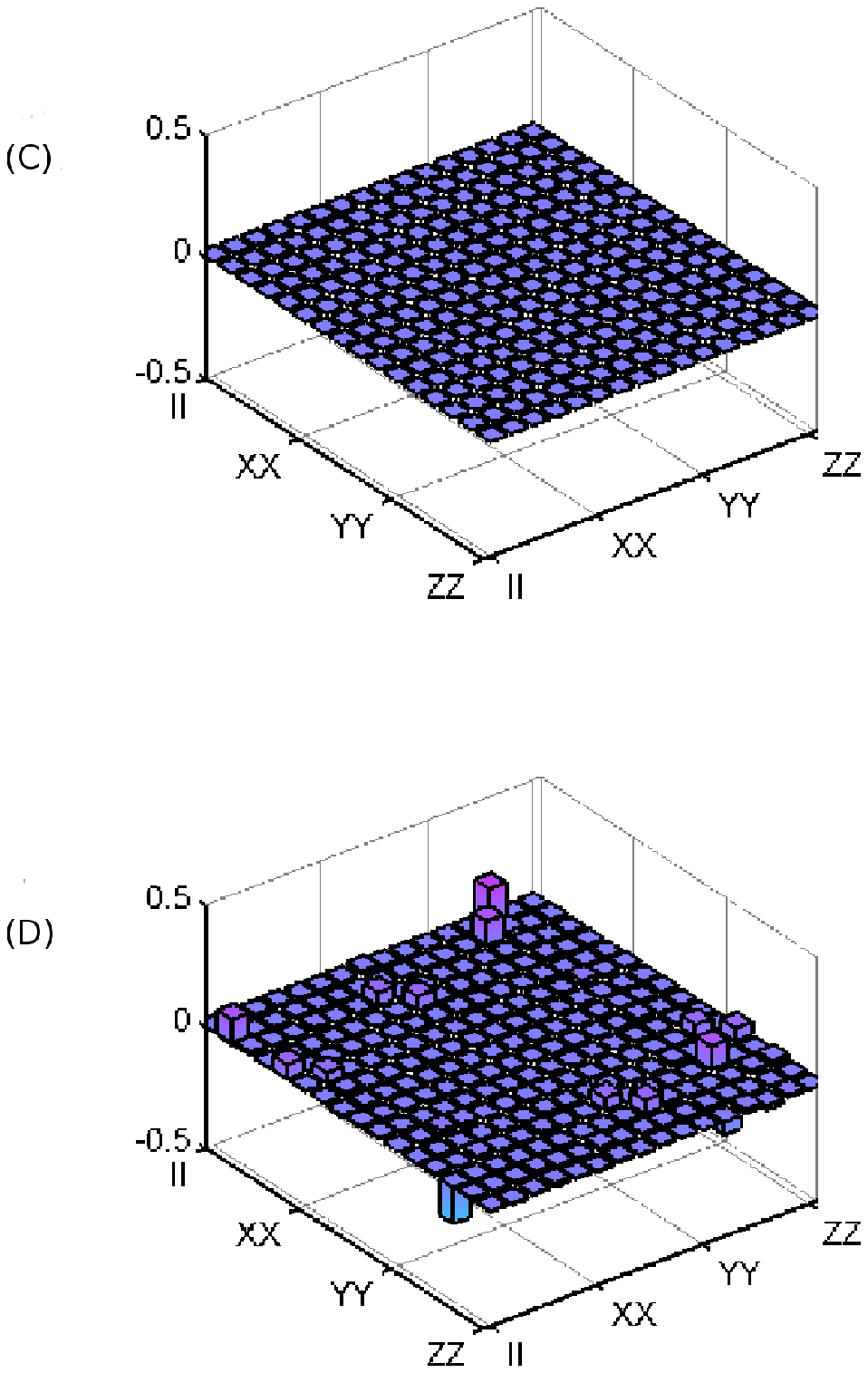}
  \caption{(A) and (B) represent real parts of ideal and experimental process matrices respectively, while (D) and (D) represent the corresponding imaginary parts, for non-local CNOT operation.}
  \end{subfigure}\par\medskip
  \begin{subfigure}{\linewidth}
  \adjincludegraphics[width=.5\linewidth]{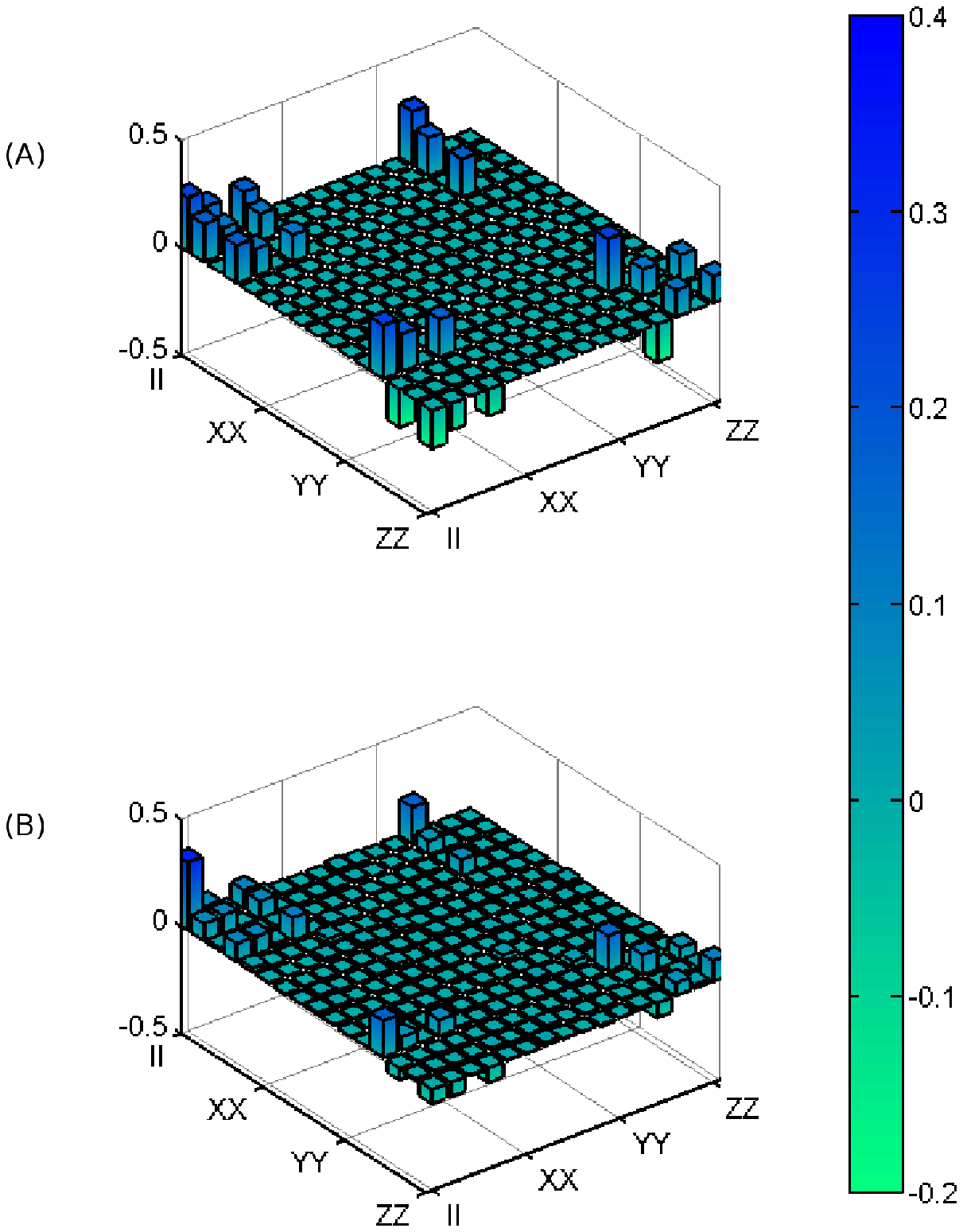}\hfill
  \adjincludegraphics[width=.5\linewidth]{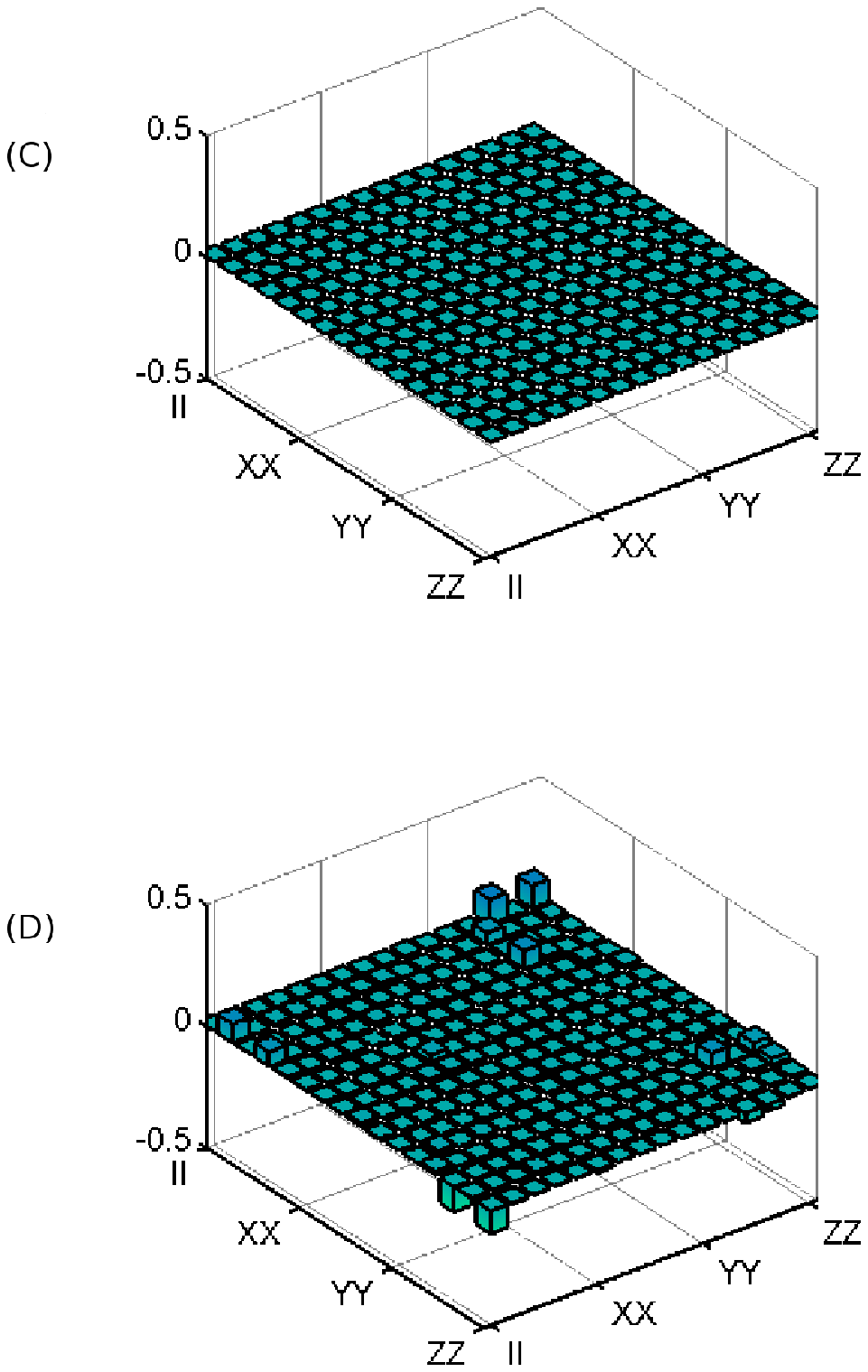}
  \caption{(A) and (B) represent real parts of ideal and experimental process matrices respectively, while (D) and (D) represent the corresponding imaginary parts, for non-local CH operation.}
  \end{subfigure}\par\medskip
  \caption{ }
  \label{fig:10}
\end{figure}

For each output state, 16 transformed density matrices $\varepsilon(\rho^{jk})$ \cite{White:07} are need to be constructed, where $j,k=1,2,3,4$. The process matrix of a two-qubit quantum operation $\chi$ can be obtained from the following relation,
\begin{equation}
    \chi=K^{T}\left[\begin{array}{cccc}
    \varepsilon(\rho^{11}) & \varepsilon(\rho^{12}) & \varepsilon(\rho^{13}) & \varepsilon(\rho^{14})\\
    \varepsilon(\rho^{21}) & \varepsilon(\rho^{22}) & \varepsilon(\rho^{23}) & \varepsilon(\rho^{24})\\
    \varepsilon(\rho^{31}) & \varepsilon(\rho^{32}) & \varepsilon(\rho^{33}) & \varepsilon(\rho^{34})\\
    \varepsilon(\rho^{41}) & \varepsilon(\rho^{42}) & \varepsilon(\rho^{43}) & \varepsilon(\rho^{44})\\
\end{array}\right]K
\end{equation}
where $K=P\Lambda$ with
$P=I\otimes[\rho^{11}+\rho^{23}+\rho^{32}+\rho^{44}]\otimes I$
and $\Lambda=(Z\otimes I+ X\otimes X)\otimes(Z\otimes I+ X\otimes X)/4$. 
Here, $\rho^{jk}$ are the matrices with elements `1' at position (j,k) and `0' elsewhere. If all input matrices can be written as a linear combination of $\rho^{jk}$, i.e.,
\begin{equation}
    \vec{\rho}^{\hspace{1ex}\alpha\beta}=M\vec{\rho}^{\hspace{1ex}jk}
\end{equation}
(where $\vec{\rho}$ represents a column matrix whose elements are the input density matrices), then the inverse of matrix $M$ can map the transformed output density matrices $\varepsilon(\rho^{jk})$ to the actual output density matrices using the relation, 
\begin{equation}
    \vec{\varepsilon}(\rho^{jk})=M^{-1}\vec{\varepsilon}(\rho^{\alpha\beta})
\end{equation}
where $\vec{\varepsilon}(\rho)$ represents a column matrix whose elements are the output density matrices. The matrix $M^{-1}$ is provided in Appendix of Ref. \cite{White:07}. If $\chi_{T}$ and $\chi_{E}$ are the trace preserving process matrices for theoretical and experimental cases respectively, then the process fidelity \cite{PhysRevLett.93.080502,PhysRevA.71.062310,PhysRevA.89.042304} is given by,
\begin{equation}
F_{p}(\chi_{T},\chi_{E}) =\left(Tr\left(\sqrt{\sqrt{\chi_{E}}\chi_{T}\sqrt{\chi_{E}}}\right)\right)^{2}=Tr(\chi_{T}\chi_{E})
\end{equation}
The above equation gives the overlap between experimental and theoretical process matrices which reduces to much simpler form in the case of trace-preserving operations and helps to determine the accuracy of action of gates on arbitrary states. In Figs. \ref{fig:10}, the ideal and experimental process matrices for the two cases have been visualized. The process fidelities for non-local CNOT and CH operations are estimated to be $F_{p}^{CN}=0.536$ and $F_{p}^{CH}=0.554$ respectively. The average gate fidelity\cite{PhysRevLett.93.080502} $\bar{F}$ is related to the process fidelity by $\bar{F}=(d.F_{p}-1)/(d+1)$, where $d=2^{N}$ for N qubit input. In our experiment, the corresponding values obtained are $\bar{F}^{CN}=0.628$ and $\bar{F}^{CH}=0.643$. This quantity represents the state fidelity between theoretical and experimental gate outputs. 
Both fidelity measurements lies within the range $0< F \leq 1$, where 1 corresponds to the ideal value.

\section{Conclusion \label{V}}
To conclude, we have explicated here non-local implementation of two different controlled unitary quantum gates, i.e., Controlled-NOT and Controlled-Hadamard gates, on IBM 5 qubit quantum computer. The techniques of quantum state and process tomographies are used to characterize the performed quantum operations. The ideal and experimental output density matrices of an arbitrary input state are compared and it is found that the Controlled-Not and Controlled-Hadamard gates are implemented non-locally with fidelities 0.879 and 0.831 respectively. The process tomography reveals the accuracy of operations of CH and CNOT gates with fidelities 0.536 and 0.554 respectively. 

\section*{Acknowledgments}
\label{acknowledgments}
VPK acknowledges IISER-K Summer Student Research Programme 2017, and INSPIRE Scholarship awarded by Department of Science and Technology, Govt. of India for support. BKB acknowledges the support of DST Inspire Fellowship. We acknowledge Haozhen Situ for helpful discussions. We express our deep gratitude to IBM team for providing access to IBM Quantum Computer. The views expressed in this work are those of the authors and do not reflect the opinions of IBM or any of its employees.

\end{document}